# The relationship between parameters and effects in transcranial ultrasonic stimulation.

*Tulika Nandi (1,2), Benjamin R. Kop (1), Kim Butts Pauly (3), Charlotte J. Stagg (4,5), Lennart Verhagen (1)*

1. *Donders Institute for Brain Cognition and Behaviour, Radboud University, Nijmegen, The Netherlands*
2. *Department of Human Movement Sciences, Vrije Universiteit Amsterdam, Amsterdam, The Netherlands*
3. *Department of Radiology, Stanford University, Stanford, CA, USA*
4. *Wellcome Centre for Integrative Neuroimaging, FMRIB, Nuffield Department of Clinical Neurosciences, University of Oxford, Oxford, UK*
5. *Medical Research Council Brain Network Dynamics Unit, Nuffield Department of Clinical Neurosciences, University of Oxford, Oxford, UK*

**Abstract**

*Transcranial ultrasonic stimulation (TUS) is rapidly gaining traction for non-invasive human neuromodulation, with a pressing need to establish protocols that maximise neuromodulatory efficacy. In this review, we aggregate and examine empirical evidence for the relationship between tunable TUS parameters and in vitro and in vivo outcomes. Based on this multiscale approach, TUS researchers can make better informed decisions about optimal parameter settings. Importantly, we also discuss the challenges involved in extrapolating results from prior empirical work to future interventions, including the translation of protocols between models and the complex interaction between TUS protocols and the brain. A synthesis of the empirical evidence suggests that larger effects will be observed at lower frequencies within the sub-MHz range, higher intensities and pressures than commonly administered thus far, and longer pulses and pulse train durations. Nevertheless, we emphasise the need for cautious interpretation of empirical data from different experimental paradigms when basing protocols on prior work as we advance towards refined TUS parameters for human neuromodulation.*

## *1. Introduction*

Transcranial ultrasonic stimulation (TUS) is an emerging tool for human neuromodulation which overcomes many of the limitations of existing invasive and non-invasive brain stimulation techniques. TUS promises high spatial resolution and access to both superficial and deep brain structures without the need for invasive procedures. Neuromodulatory effects of TUS have been demonstrated both *in vitro* and *in vivo* in a range of species including worms, rodents, sheep, non-human primates and humans[1]. In parallel, mechanistic models have been developed to elucidate the underlying biophysical and neurobiological mechanisms through which ultrasound interacts with neurons and glial cells.

Currently, a major challenge for the application of TUS is the development and selection of protocols which maximise the desired effects for each experimental or clinical goal by appropriately tuning sonication parameters. Ideally, protocols would be designed based on theoretical knowledge of ultrasound biophysical effects, their interactions with neurophysiology and their relationships with stimulation parameters. A more practical, though resource-intensive, approach is to empirically compare parameters and make *post hoc* inferences about the biophysical mechanisms of TUS based on the known relationships between different mechanisms and tunable parameters.

In this review we have compiled empirical research on tunable parameters across model systems and into humans, with the aim of aiding TUS users in making informed decisions about parameter settings (see Tables 1-3). We begin by discussing the value and challenges of a multiscale approach i.e., studying ultrasound effects from the cellular to clinical levels in parallel. Next, we consider the interactions between parameters or protocols, and the brain itself. Then, we briefly describe the relationships between ultrasound biophysical effects and tunable parameters (for a detailed discussion please see[2]). Finally, we summarise the currently available empirical evidence regarding the effects of manipulating fundamental frequency ($f_0$), amplitude, and pulsing parameters.

## *2. Measuring ultrasound neuromodulatory effects: a parallel multiscale approach*

Many different *in vitro* substrates and *in vivo* (Figure 1) models, along with a wide range of outcome measures have been used to examine the effects of tuning ultrasound parameters. This parallel, multiscale approach is crucial for eventually establishing a continuous chain of evidence that allows us to fully elucidate the mechanisms underlying TUS and hence justify parameter choices, for experiments ranging from those at the cell membrane level to behavioural and clinical studies. Results from each level of explanation can inform and help design studies at other levels, thereby accelerating progress in the field, even at a stage when our understanding at each level is incomplete. One fitting example of the benefits of a multiscale approach is the successful suppression of spike activity by TUS in a mouse model of chronic temporal lobe epilepsy using a protocol previously discovered to effectively increase hippocampal inhibitory interneuron activity in awake mice[3]. In parallel, clinical pilots have demonstrated preliminary safety[4] and efficacy[5] of TUS for reducing seizure frequency in human epilepsy patients, and follow-up trials can benefit from the knowledge of protocols gained from animal work.

However, this multiscale approach also presents some challenges. For instance, *in vivo* outcomes like EMG that represent population-level suprathreshold effects might demonstrate a different parameter dependence compared to *in vitro* single neuron effects, or subthreshold effects that might activate or favour a distinct population of neurons. This issue is particularly relevant since human studies have so far focused on subthreshold effects, while animal work has often used suprathreshold outcomes, making it difficult to extrapolate from animal findings to humans.

Different model systems pose different challenges, meaning that individual TUS protocols or bioeffects are more tractable in some systems than others. Therefore, direct translation from one level to another is not always trivial. For instance, patch clamp set-ups can be unstable at the low frequencies (sub-Megahertz range) which are used for *in vivo* transcranial applications[6,7]. Therefore several *in vitro* studies have used a high $f_0$ (43 MHz)[8,9] which is likely susceptible to fewer technical challenges but is outside the plausible range for transcranial applications in humans due to skull attenuation.

Similarly, pulsed TUS protocols with a sharp onset and offset are often used *in vitro* but lead to auditory co-stimulation *in vivo*. This is a substantial experimental issue as, in addition to preventing successful blinding, recent data suggest that the sound can directly influence corticospinal excitability which has been used as an outcome measure in a number of studies[10]. In animal studies, auditory co-stimulation can influence neuronal spiking activity[11] and calcium signalling[12], and even lead to a startle response which would confound EMG and movement measurements, the most commonly used outcome measures to date. In animal models, additional controls such as deafening are feasible and can be used to mitigate auditory co-stimulation[11,12]. Indeed, later work using protocols that elicited relatively small or no auditory brainstem responses, demonstrated motor responses to TUS in both hearing and deaf mice[13]. If we want to translate these animal protocols, and other effective but audible *in vivo* protocols to humans, it will be essential to either carefully choose outcome measures that are not influenced by the auditory co-stimulation, or use appropriate control and masking measures to ensure that any observed effects can be attributed to the stimulation itself.

A parallel, multiscale approach is essential for an in-depth understanding of TUS effects. However, this approach will only succeed when parameters and effects can be translated across scales. It is necessary to consider the challenges discussed above when critically evaluating the available TUS data (presented in Tables 1, 2 and 3). To this end, we include a summary of specific controls used in each study in the tables. When planning a TUS study using protocols previously reported in the literature, we urge readers to carefully consider the choice of TUS protocol in relation to their chosen model (*in vitro*, *in vivo,* or human) and outcome measure.

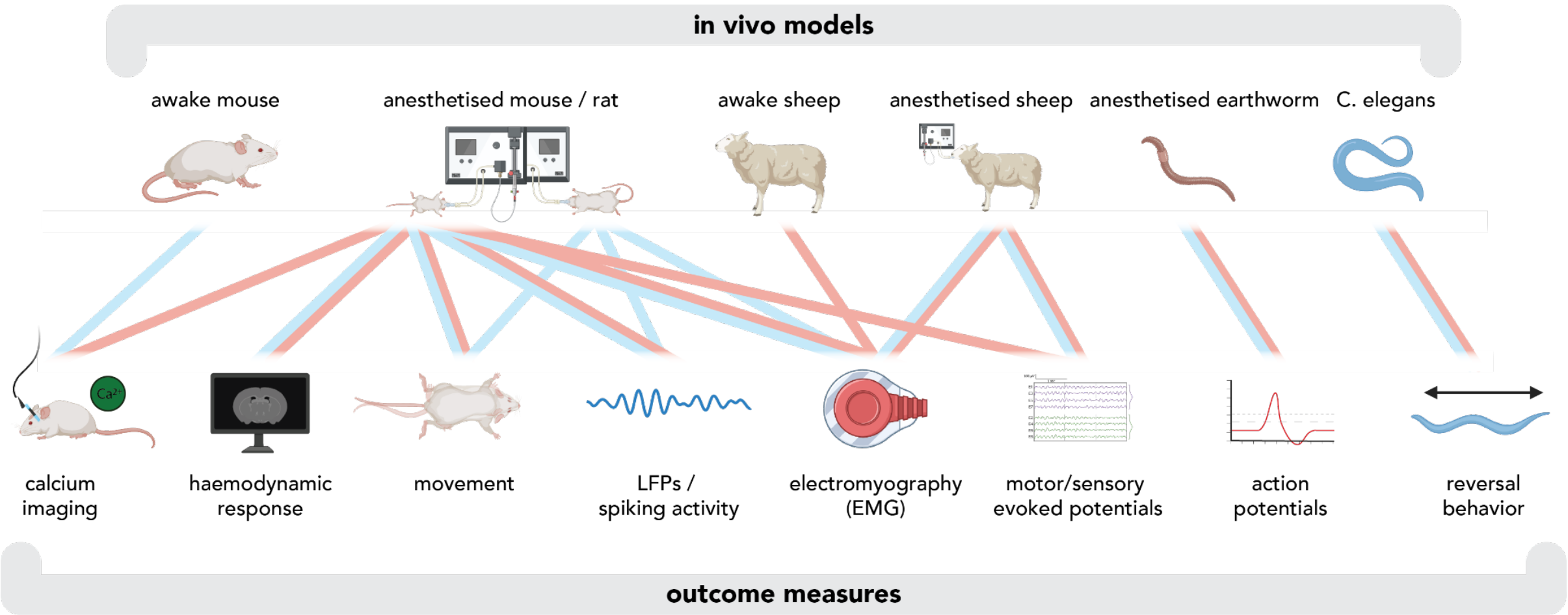

**Figure 1.** In vivo models and outcome measures used in studies examining tuning ultrasound parameters, including research investigating pressure/intensity (blue) and pulse/pulse train duration (red). Created with BioRender.com.

### *3. Brain-protocol interactions*

The ultimate effect of any TUS protocol is determined by interactions between the parameters, and the static and dynamic characteristics of the brain. Though this review will focus on protocol parameters, it is essential to note that any description of the likely effects of various TUS parameters is incomplete without knowledge of the neural target. Important considerations are the level of description (e.g. membrane, circuit, or behaviour), region-specific properties, temporal dynamics including state-dependency, and changes over lifespan or between health and disease.

First, TUS effects are not limited to one level of description but cross from biophysics to cellular biomechanisms, to circuit-level neurophysiology, to the human brain and behaviour, and to clinical outcomes. Critically, ultrasound, therefore, can have different, even opposing, effects on individual levels. For example, an ultrasound protocol might facilitate calcium influx and spiking activity at the membrane level, exciting the stimulated neurons[14]. Simultaneously, some pulsing protocols have been shown to selectively stimulate GABAergic inhibitory neurons, leading to circuit-level inhibition[3]. Such a protocol has both excitatory and inhibitory effects, depending on the level of description. Similarly, a computational model, the neuronal intramembrane cavitation excitation (NICE) model, also predicts that different protocols preferentially activate excitatory or inhibitory neurons, and the net effect would be determined by the proportion of different types of neurons in the target region[15,16]. In summary, it is therefore critical to consider the level of organisation when describing the effects of an intervention.

Spatially, the same TUS protocol may elicit different effects in different brain regions. The effects of TUS will depend on several factors including the relative proportion of inhibitory versus excitatory neurons, or concentrations of different neurotransmitters. For instance, in mice, the same ultrasound protocol elicited short bursts of activity in the primary motor cortex, and longer-lasting rhythmic bursting in the hippocampus[6], demonstrating the interaction between ultrasound and the intrinsic characteristics of the targeted brain region. Additionally, the mechanical properties of the target neural tissues will impact the neuromodulatory effects of ultrasound. Both at macroscopic (whole brain) and microscopic levels (cell membrane), neural tissues demonstrate viscoelastic properties[17]. This implies that, given a fixed acoustic pressure or ARF, the magnitude of the resultant strain will depend on material properties (like stiffness) of the target tissues[18], which might vary across brain regions.

Temporally, the state-dependence of NIBS effects is well known[19,20], and has also been demonstrated for ultrasound. The same protocol has been shown to increase or decrease neuronal firing frequency depending on the underlying neuronal activity at the time of ultrasound application[21–23]. In addition to intrinsic factors, experimental factors like the level of anaesthesia are also known to influence ultrasound effects[24,25]. So far, the majority of *in vivo* studies have examined the effects of parameter modulation in lightly anaesthetised animals, and further work is necessary to confirm whether these findings can be replicated in unanesthetized animals. Finally, any target brain region is part of a larger network, and the effects of any protocol must be viewed in light of the neurophysiological characteristics and temporal state of the entire network. The NICE model mentioned earlier also predicts that the net ultrasonic neuromodulatory effect measured in a cortical region is modulated by the strength of the thalamic inputs to the cortex during sonication.

Lastly, these effects will not only vary between regions and brain states, but also across the lifespan and between health and disease. For example, even fundamental mechanical properties of the brain change due to ageing[26] and pathology[27].

In summary, the neuromodulatory effect of a protocol is not a property of the stimulation protocol itself, but of the interaction between the protocol and the stimulated neural circuit.

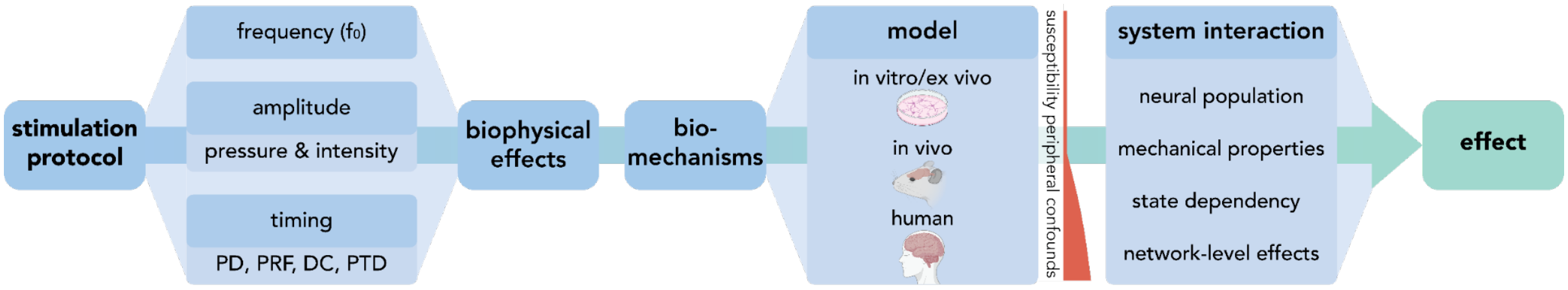

**Figure 2.** Relationship between the stimulation protocol and the net effect of TUS. Properties of the stimulation protocol, including the fundamental frequency ($f_0$), pressure/intensity, and temporal characteristics, result in a profile of biophysical effects. These effects interact with the biomechanisms in the stimulated model, whether *in vitro*/*ex vivo*, *in vivo*, or human. Each model may respond differently to the same protocol and have varying susceptibility to peripheral confounds. The stimulation effects further depend on the stimulated system, including factors such as neural population/brain region, mechanical properties of that region, the state of the system, and potential network-level interactions. Finally, the net effect of the stimulation protocol is elicited. When evaluating the effects of stimulation protocols we must consider these intermediate processes. Icons created with BioRender.com.

## *4. Biophysical effects of ultrasound and their relationships with tunable parameters*

Ultrasound interacts with tissues through mechanical forces, likely via multiple mechanisms simultaneously. Possible mechanisms include particle displacement strain, acoustic radiation force (ARF) strain, and acoustic cavitation[2]. The relative contributions of these different mechanisms are a function of the applied stimulation parameters. For example, a higher $f_0$ results in greater ARF strain, while particle displacement strain is independent of $f_0$. Particle displacement itself (not strain), on the other hand, is higher at low $f_0$. This knowledge has been exploited to develop hypotheses about the primary biophysical mechanism driving neuromodulatory effects. For instance, one study[28] found that when ultrasound is applied peripherally, a lower $f_0$ is more effective at eliciting sensations, including pain which is mediated by unmyelinated nerve endings and might therefore have similarities with neurons in the brain. This result provides empirical support for particle displacement as ultrasound neuromodulation's primary effective biophysical component, at least within the given experimental context. However, as discussed in later sections, different studies suggest different biophysical effects as the primary effective component and currently there is no consensus in the literature.

In the following sections, we will discuss multi-scale empirically observed relationships between tunable parameters and neuromodulatory outcomes. These parameters include $f_0$, amplitude characteristics: pressure and intensity, and temporal characteristics: pulse duration and pulse repetition frequency. Tables 1, 2, and 3 provide an overview of *in vitro/ex vivo, in vivo,* and human studies respectively.

## *5. Stimulation parameters and neuromodulatory outcomes*

### *5.1. Fundamental frequency*

Fundamental frequencies ranging from approximately 200 kHz to several tens of MHz have been used in studies examining the neuromodulatory effects of ultrasound.

For *in vitro* studies*,* higher frequencies seem to be more effective for eliciting action potentials[8] and calcium responses[14] than lower frequencies In the megahertz range, neuromodulatory effects may be driven by the ARF strain or acoustic streaming[7,8]. ARF is proportional to the $f_0$ and intensity, and since higher frequencies are more strongly attenuated by the skull[29], skull heating would likely limit the intensities, and consequently the ARF strain, which can be achieved transcranially *in vivo*. This is especially limiting in larger animals, with a relatively thick skull, including humans. Additionally, though acoustic streaming has been observed *in vitro*[7], the physical constraints experienced by in vivo neurons embedded in an extracellular matrix are very different from the physical constraints *in vitro*. Therefore, it is unclear whether *in vivo* neurons would experience similar mechanical effects from acoustic streaming.

*In vivo* studies suggest that lower frequencies (in the kHz range) are more effective than higher frequencies (high kHz or MHz range) for eliciting action potentials[30], EMG responses[25,31] and measurable movements[32]. While some of these results might be explained by skull attenuation, this $f_0$ dependence is also observed in earthworms[30] and mice[25], where attenuation is either irrelevant or negligible. One potential explanation is that the larger focal area at low frequencies leads to a larger volume of neural tissue being stimulated. However, even when equalising focal volume for different fundamental frequencies, studies have suggested that lower $f_0$ is more effective for *in vivo* neuromodulation[33,34]. Indeed, we cannot rule out that neuromodulation is driven by particle displacement or cavitation-based mechanisms when TUS is applied in the kilohertz range. A computational model[15], based on membrane deformation driven by particle displacement strain, predicts higher efficacy at lower frequencies, possibly because the mechanical properties of cell membranes limit the deformations achieved at higher frequencies. These observations do not necessarily contradict the *in vitro* findings but rather highlight the possibility that different underlying mechanisms might drive ultrasonic neuromodulatory effects in different situations.

Overall, in light of the current evidence, and the potential for skull heating at higher frequencies, sub-MHz frequencies seem to be most appropriate in humans.

### *5.2. Pressure and Intensity*

If all other parameters are constant, increasing TUS pressure increases particle displacement strain, ARF strain, acoustic streaming, the probability of cavitation, and heating.

Therefore, TUS amplitude over time is an important factor in neuromodulation, irrespective of the primary mechanism of action. Indeed, multiple studies have found larger effects when increasing pressure/intensity (see Tables 1-3). However, the exact nature of the dose-response relationship is as yet unclear. Specifically, while some data suggest that a minimum amplitude is required to elicit any effects[8], the threshold has not been identified. In practice, the threshold is likely to vary for different types of neurons, brain targets and intended outcomes. Identification of intensities at which neuromodulatory TUS effects plateau or even potentially decrease[35,36], will help to determine the effective range of intensities in humans and minimise any unintentional side effects. The majority of studies to date suggest higher TUS amplitudes result in greater neuromodulatory effects, indicating that the intensity range used is below any potential ceiling. Therefore, where feasible, it would be reasonable to increase amplitude in future work.

An important consideration when investigating the effects of pressure/intensity on neuromodulatory outcomes is the potential for dose-response relationships of peripheral confounds. For example, several *in vivo* studies have examined the effects of ultrasound parameters on EMG or movement. One concern is that the EMG or movement response is a startle elicited by the audible sound, and could scale with loudness i.e., pressure/intensity. In some studies, a startle response can be distinguished from direct neuromodulatory effects using various criteria. For instance, a relatively long latency[13,25] (e.g., after 80ms rather than the 6-8ms latency commonly observed for the acoustic startle response commonly in rodents[37]), or the scaling of response duration with stimulus duration rules out a startle response[13]. In humans, volume scales with intensity, and cueing effects therefore might also scale with intensity[10]. Since not all studies include adequate outcome information or controls to rule out confounding effects, and any reported dose-response effects must be interpreted with caution.

TUS intensity will spatially vary within the ultrasound focus, and the intensity does not immediately drop to zero beyond the commonly reported -6dB or -3dB boundary of the focus. Therefore, neurons at different spatial locations will experience different doses and any functional outcomes will depend on the spatially cumulative effects. Finally, depending on the relevant biophysical effect of ultrasound, neuromodulatory effects may be correlated more strongly with either pressure or intensity. Identification of the crucial parameter (pressure or intensity) and knowledge of the nature of the relationship (linear, quadratic or other) will allow for more precise dose titration.

### *5.3. Pulse repetition frequency and pulse duration/duty cycle*

The majority of human TUS studies so far have used pulsed ultrasound, and some data[38] and theoretical models[15] suggest that pulsed ultrasound is more effective than continuous at a given intensity. One explanation is that, akin to transcranial alternating current stimulation (tACS) or rTMS, the pulse repetition frequency mimics or interacts with biologically relevant frequencies. However, in the peripheral nervous system, it has been shown that pulsed ultrasound is more likely to cause tactile sensations than continuous ultrasound[34], and possible somatosensory confounding effects cannot be ruled out. Additionally, though some preliminary data suggest that higher PRFs are more effective[25,38–40], the auditory confound was largely overlooked in this early work. Rodents are more sensitive to the higher frequencies[41] in the range of tested PRFs (30 to 10,000 Hz) and higher efficacy at these PRFs may be driven by higher audibility, rather than true neuromodulatory effects. In favour of frequency dependence, one *in vitro* study, where sensory confounds are irrelevant, found a PRF of 1500 Hz to be more effective than 300 Hz[39]. Additionally, one *in vivo* study demonstrated PRF dependence only in excitatory and not in inhibitory neurons, suggesting that PRF effects may be complex and non-linear[40]. In both of these studies, the total energy delivered was matched across different PRFs by adjusting the PD.

The studies discussed so far examined the immediate, online effects of ultrasound. In offline studies examining prolonged effects outlasting the stimulation itself, both 10 Hz[42–46] and 5 Hz[47,48] PRFs are effective across multiple non-human primate and human studies. However, there is very limited data directly comparing[47], and demonstrating the superiority of these PRFs over others. Overall, the manipulation of PRF to enhance TUS effects is still a promising line of enquiry, especially when it is informed by the potential biological relevance of specific frequencies.

So far, we have focused on the effect of PRF in studies which have matched energy deposition across different PRFs. Alternatively, at a chosen PRF, total energy deposition can be altered by changing the duty cycle (DC) or PD. We hypothesise that increasing the DC, and hence increasing dose, would increase the efficacy of ultrasound in a manner analogous to continuous applications. For example, prolonged TUS exposure could either lead to cumulative effects on a single neuron, potentially by allowing greater time for ion movement and changes in membrane potential[14], or increase the probability of recruiting additional neurons. While there is some preliminary evidence to support this hypothesis[30], other studies have either found no effect of increasing DC[49], or have simultaneously manipulated several parameters[32,50,51], making it difficult to infer any effects specific to the DC. In theory, a PRF-dependent minimum DC would be necessary to ensure that the pulse duration exceeds a threshold required for any meaningful interaction with neurons, but this threshold has yet to be determined. It is also unclear how far the DC can be increased to obtain a dose benefit without losing any potential benefits of pulsing.

An alternative hypothesis might be that rather than simply altering the dose, the proportion of ultrasound on versus off time is crucial for interactions with neurons. One computational model[16] suggests that low DCs preferentially activate inhibitory neurons while higher DCs activate both inhibitory and excitatory neurons. In this model, the differential TUS effect on inhibitory and excitatory neurons is driven by the properties of specific ion channels found predominantly on inhibitory neurons. While there is some preliminary evidence of net inhibitory effects at low DC and *vice versa*[51], the different TUS protocols were not compared using the same outcome measure. Instead, different outcome measures were chosen specifically to detect excitatory versus inhibitory effects[50]. Therefore, further systematic investigation is necessary to understand the effects of manipulating DC in pulsed protocols.

So far, we have only discussed one layer of temporal patterns, from pulses to pulse trains. However, pulse trains themselves may then be repeated at different frequencies[52], creating a second order temporal pattern. Such nested temporal patterns are known to be relevant for various biological processes and are commonly used in other neuromodulatory techniques like TMS. This might also be relevant to TUS but has not been explored extensively yet.

### *5.4. Interaction between amplitude and temporal characteristics*

Both *in vitro* and *in vivo* data, including early human results, show a scaling of outcomes with both intensity and duration of sonication (see Tables 1-3). Preliminary data also suggest that, analogous to the strength-duration relationship for electrical stimulation, a higher intensity is required to elicit an outcome with shorter sonications and vice-versa[8,30]. Along these lines, *in silico* data[53] examining neuronal membrane deflection and consequent change in capacitance, suggest that a minimum sonication duration is necessary to trigger action potentials, with longer durations required at lower pressures. Indeed, it is likely that there is a minimum pressure or duration necessary to elicit a response when using stimuli that are very long or very high pressure respectively. Systematic *in vitro* and animal work to identify thresholds is crucial to avoid underdosing in human and clinical studies.

The combination of amplitude and the time over which it is administered can also be described using dose-rate[2,54]. To the best of our knowledge, there is no empirical data evaluating the effects of manipulating dose-rate.

## *Conclusions*

Taken together, the studies reviewed here provide compelling evidence for neuromodulation across multiple levels of organisation, and provide a basis for evidence based parameter selection. The currently available empirical data favour the use of relatively low fundamental frequencies (kilohertz range), higher intensities and longer pulse or pulse train durations for TUS in humans. Theoretically, the effects might plateau or even decrease at higher intensities and durations. However, this upper limit has yet to be identified, and given that human studies to date have predominantly demonstrated sub-threshold effects, it is reasonable to investigate the effects of increasing intensities and durations, within safe limits[55]. There is preliminary evidence that pulsing can be valuable, and in some situations, may elicit stronger effects compared to continuous protocols, even when the total energy deposition is matched. However, the optimal frequencies for pulsing remain unknown, and further work is required to test whether ultrasound can entrain, enhance or otherwise interact with biologically relevant frequencies[56]. Additionally, it is unclear whether manipulation of the duty cycle simply alters the dose, or has additional value due to the sensitivity of ion channels and neurons to the proportion of ultrasound on versus off time within each repetition of a pulse[16]. The *in vitro* substrates, animal models and outcome measures employed in empirical studies are extremely variable. Additionally, they cover only a small subset of the vast multi-dimensional parameter space. Therefore, while some common patterns emerge, we must also rely on theoretical knowledge about ultrasound biophysical effects, not only when choosing protocols for various applications, but also to specifically design studies for optimising parameter selection.

**Table 1:**
Summary of *in vitro* and *ex vivo* studies that directly compare TUS parameters.

| | ***Substrate*** | ***Outcome measure(s)*** | ***Directly compared TUS parameters*** | ***Relevant findings*** | ***Shams/ controls in parameter comparison experiments*** | ***Notes*** |
|---|---|---|---|---|---|---|
| *Sorum* et al., *2023*[57] | Xenopus laevis oocytes expressing human TRAAK channels | Current (measured using inside-out patch clamp) | $I_{sppa}$ | TRAAK and MscS current increases with increasing $I_{sppa}$. | Both TRAAK and MscS channels have similar low activation thresholds, but TRAAK shows a broader response range. | PW |
| *Yoo* et al., *2022*[14] | Primary mouse cortical neuron cultures | Calcium imaging | $f_0$ (see supplement), $I_{sppa}$ and PD | Response magnitude is higher at 670 kHz compared to 350 kHz. Response magnitude increases with increasing $I_{sppa}$ or PD. Response latency decreases with increasing intensity. | Response magnitude is larger and increases with increasing $I_{sppa}$ when mechanosensitive channels are overexpressed. | CW |
| *Weinreb and Moses, 2022*[58] | (Pharmacologically disconnected) Rat hippocampal neural cultures | Calcium imaging | Pressure and PD | More cells are activated and latency is longer with 40 ms pulses compared to 4 µs pulses. More cells are activated at 0.67 MPa compared to 0.35 MPa. Pressure dependence measured in presence of P2 purinergic receptor blocker. | None. | CW |
| *Fan et al. 2022*[59] | Rat cortical neural and glial cultures | Calcium imaging | $f_0$ and $I_{sppa}$ | Higher $I_{sppa}$ required to elicit responses at 500 kHz compared to 39.5 kHz. At both frequencies, response magnitude increases, and | None. | CW |
| | | | | response latency decreases, with increasing $I_{sppa}$. | | |
| *Suarez-Castellanos* et al.*, 2021*[60] | Mouse hippocampal slices | Field excitatory post-synaptic potentials (fEPSP; measured using multi-electrode array) | Pressure | fEPSP amplitude and slope increases, and latency decreases, with increasing pressure. | Recording from electrode away from sonicated site. LFPs recorded from the control (unstimulated) site only at the highest focal pressure. | PW |
| *Manuel* et al.*, 2020*[39] | Coronal brain slices containing motor cortex from transgenic mice selectively expressing genetically encoded calcium indicators in cortical pyramidal cells or all neuronal cells | Calcium imaging | PRF | Change in calcium signals (relative to baseline) with 1500 Hz but not with 300 Hz PRF. | None. | PW<br>Total energy delivered is matched by decreasing PD at the higher PRF. |
| *Menz* et al.*, 2019*[8] | Isolated salamander retina | Ganglion cell spiking | $f_0$, $I_{sppa}$ and PD | Higher $I_{sppa}$ threshold for eliciting response using lower $f_0$ and shorter PDs. Firing rate increases with increasing $I_{sppa}$. Predictions of in-silico model based on an ARF driven mechanism match empirical $f_0$ and $I_{sppa}$ dependence data. | None. | CW |
| *Qiu* et al.*, 2019*[61] | HEK293T cells with heterologously overexpressed mouse Piezo1 or mouse primary neuron cultures with endogenously expressed Piezo1 | Calcium imaging | Pressure | Response magnitude increases with increasing pressure in transfected HEK293T cells. Response magnitude increases with increasing pressure in mouse neuron cultures. | Pressure dependence not seen in genetic control without heterologously expressed Piezo1. | PW |
| *Menz* et al., *2013*[9] | Isolated salamander retina | Ganglion cell spiking | $I_{spta}$ and PRF | Firing rate and response latency increase with increasing $I_{spta}$. No effect of modulating PRF in the 15 Hz to 1 MHz range. | None. | CW and PW |
| *Tyler* et al., *2008*[35] | Mouse hippocampal slice cultures | Synaptic vesicle exocytosis | $I_{spta}$ | With increasing $I_{spta}$, synaptic vesicle exocytosis increases initially and then decreases. | None. | PW |

*$I_{sppa}$: spatial peak pulse averaged intensity; $I_{spta}$: spatial peak time averaged intensity; $f_0$: fundamental frequency; PD: pulse duration; PRF: pulse repetition frequency; CW: continuous wave; PW: pulsed wave*

**Table 2:**
Summary of *in vivo* studies that directly compare TUS parameters.

| | ***Animal model*** | ***Outcome measure(s)*** | ***Directly compared TUS parameters*** | ***Relevant findings*** | ***Shams/ controls in parameter comparison experiments*** | ***Notes*** |
|---|---|---|---|---|---|---|
| *Kim et al. 2024*[52] | Anesthetised mice | Ultrasound evoked-MEP and motor skill acquisition | Pulse train duration, pulse train repetition frequency, pulse train repetition duration, pulse train repeat repetition duration (only applies to iTBUS) and number/ pattern of nested temporal layers. | MEP amplitude increases after iTBUS and decreases after cTBUS. For both protocols, a larger number of total pulses, i.e., a longer pulse train repeat duration, resulted in longer lasting effects. iTBUS was effective at a 5 Hz pulse train repetition frequency, but not at 10 Hz. iTBUS (with gamma bursts) was more effective than i-Theta (no gamma bursts). For iTBUS, 4 repeats of 30 Hz pulse trains (bursts) within the 5 Hz pulse train repetition frequency yielded stronger effects compared to 3, 5, or 6 repeats. Motor skill acquisition success rate was higher for iTBUS compared to cTBUS.<br><br>iTBUS and cTBUS were also compared using several measures which provide further information about the biomechanisms underlying the MEP effects. | Inactive sham with no ultrasound delivered.<br><br>Motor learning: active control site | PW<br>Offline |
| *Zhu* et al., *2023*[62] | Anesthetised mice | Limb movement distance, EMG amplitude and calcium imaging (using fibre photometry) | Pressure and pulse train duration | EMG amplitude and calcium response magnitude increases with increasing pressure. Movement duration increases with increasing pulse train duration. | Stronger ultrasound effects in controls compared to Piezo1 knockout mice. | PW |
| *Di Ianni et al. 2023*[63] | Awake rats | Arousal quantified by speed and distance travelled. | Intensity and number of sonications | A single PTD of 5 seconds (PD = 80 ms, PRI = 480 ms) did not have any effects, while three pulse trains with a 10 second inter-train interval did have an effect on both speed and distance travelled for some intensities. Here, an inverted U-shaped dose-response effect was observed. | Speed and distance travelled when an active control target was stimulated. | PW |
| *Yang et al. 2022*[64] | Anesthetised macaque | BOLD signal measured using fMRI | Pressure | At rest, BOLD signal at the target somatosensory cortex, and several other areas which are known to be part of the tactile network, increases with increasing pressure. When TUS is applied concurrently with tactile stimulation, the dose-response curve varies between brain regions. | BOLD signals in control areas (auditory cortex and mediodorsal nucleus) are weak and do not show pressure dependence. | PW |
| *Vion-Bailly* et al.*, 2022*[30] | Anesthetised earthworms (Lumbricus terrestris) | Success rate of eliciting action potentials | $f_0$, $I_{sapa}$, $I_{sata}$, PD, pulse train duration and PRF | Success rate increases with increasing $I_{sapa}$, PD and pulse train duration. Higher success rate at 125 Hz compared to 25 Hz PRF, and 1.1 MHz compared to 3.3 MHz $f_0$. Success rate increases with increasing $I_{sata}$, irrespective of whether the increase is achieved by increasing $I_{sapa}$ or PD. | None. | PW<br>In some experiments, multiple parameters changed simultaneously. |
| *Murphy* et al.*, 2022*[3] | Awake mice | Calcium imaging (using fibre photometry) | Pressure | Magnitude of calcium response increases with increasing pressure. | Transducer moved while acquiring calcium signals from the same location. Smaller calcium response with off-target compared to on-target stimulation. | CW |
| *Kim et al. 2022*[65] | Awake sheep | EMG success rate, amplitude and onset latency | PD and PRF | No effect of PD and PRF on EMG success rate. Higher EMG amplitudes and shorter latencies with shorter pulses i.e., higher PRFs. | Two stimulation sites: shorter onset latency for thalamic compared to M1 stimulation. | CW<br>Total energy delivered is matched by |
| | | | | | Lateralized responses observed. | increasing PRF as PD decreases. |
| *Yuan et al. 2021*[66] | Anesthetised mice | Amplitude and power of sharp wave ripples in local field potentials and haemodynamic response measured using optical imaging | Pressure | Sharp wave ripple amplitude and power, and haemodynamic response amplitude increases with increasing pressure. | None. | PW<br>Responses to ultrasound observed only at low anaesthesia levels. |
| *Yu* et al.*, 2021*[40] | Anesthetised rats | Multi-unit activity, specifically spiking rate, measured using implanted extracellular electrodes | PRF | Spiking rate of excitatory neurons increases with increasing PRF but that of inhibitory neurons is not influenced by PRF. No effect of PRF in sham conditions. | Transducer flipped sham and control site with stimulation of bone. | PW<br>Total energy delivered is matched by decreasing the PD as PRF increases. |
| *Liu et al. 2021*[67] | Anesthetised rats | Local field potentials (measured using implanted electrodes) | Intensity and PRF | 500 and 1000 Hz PRFs are linked to gamma band activity, and 1500 Hz theta and delta band activity. An inverted U-shape relationship is observed between intensity and the number of responding nuclei. | None. | PW |
| *Yuan et al., 2020*[49] | Anesthetised mice | Cortical hemodynamic response (measured using laser speckle contrast imaging) | $I_{sppa}$, pulse train duration and DC | Haemodynamic response magnitude increases with increasing $I_{sppa}$ and pulse train duration. No effect of DC. | None. | PW |
| *Yoon* et al*., 2019*[50] | Anesthetised sheep | EMG success rate and somatosensory evoked potentials (SEP) | $I_{sppa}$, PD and DC | Higher EMG success rate with lower $I_{sppa}$, lower PD, and 70% DC compared to 30 and 100%. Some protocols lead to relatively higher suppression of SEPs, but individual parameters are not systematically compared. | Active control site in thalamus and no-US sham. Thalamic stimulation leads to EMG responses but with no $I_{sppa}$ dependence. Higher rates of contralateral EMG response after both M1 and thalamus stimulation. No effect of thalamus or ipsilateral S1 sonication on SEP. | CW and PW Multiple parameters manipulated simultaneously. |
| *Wang et al. 2019*[68] | Anesthetised mice | Power and entropy of LFPs measured using implanted microelectrodes | $I_{sppa}$, pulse train duration and DC. | With increasing $I_{sppa}$, theta power decreases, gamma power increases and entropy decreases. With increasing pulse train duration, gamma power increases and entropy decreases. All changes are limited to a few post-sonication time bins with no parameter effects observed after 1s. No effect of DC. | None. | PW |
| *Mohammadjavadi* et al*., 2019*[13] | Anesthetised mice | EMG success rate and duration | $I_{sppa}$ and pulse train duration. | EMG success rate increases with increasing $I_{sppa}$. EMG duration increases with increasing pulse train duration. No effect of pulse train duration on EMG latency. | Deaf mice: intensity response similar to hearing mice. Low (likely chance) response in sham no-US condition. | CW |
| *Kubanek* et al*., 2018*[38] | C. elegans | Reversal behaviour | Pressure, pulse train duration, DC and PRF | Reversal frequency increases with increasing pressure and pulse train duration. Reversal frequency increases with increasing PRF up to 1 kHz, and then decreases slightly at 3 and 10 kHz. Reversal frequency has an inverse U-shaped association with DC, with the | Pressure dependence is abolished in mechanosensation-defective mutants but not in thermosensation-deficient mutants. | PW Total energy delivered is matched between PRFs by changing PD, but increases |
| | | | | highest response frequency seen at 50% DC. | | with increasing DC. |
| *Kim, Anguluan and Kim, 2017*[69] | Awake mice | Cortical hemodynamic response (measured using optical intrinsic signal imaging) | PRF | Amplitude of change in haemodynamics increases with increasing PRF. | No haemodynamic changes observed in no-US sham condition. | PW<br>Total energy delivered is matched by decreasing PD as PRF increases. |
| *Ye, Brown and Pauly, 2016*[31] | Anesthetised mice | EMG success rate | $f_0$ and $I_{sppa}$ | EMG success rate increases as $I_{sppa}$ increases and is higher at lower $f_0$. Higher efficacy at lower $f_0$ is not likely to be due to larger focal spot. | Low (likely chance) response in sham no-US condition. | CW |
| *Kim et al., 2015*[51] | Anesthetised rats | Visual evoked potentials (VEP) | $I_{sppa}$ and DC | VEP amplitude is decreased at 3 W/cm$^2$ and increased at 5 W/cm$^2$, with no effects at 1 W/cm$^2$. VEP amplitude is decreased at 5% DC and increased at 8.3%, with no effect at 1%. | No effect of no-US sham on VEP amplitude. | PW<br>Total energy delivered is matched by increasing PRF as DC increases. |
| *Kim* et al.*, 2014*[32] | Anesthetised rats | Tail movement detected using motion sensor | $f_0$, PD, pulse train duration and DC | Threshold $I_{sppa}$ for eliciting tail movement is lower at 350 kHZ compared to 650 kHz $f_0$. In general, threshold $I_{sppa}$ for eliciting tail movement is higher for 30% DC compared to 50% and 70% and threshold $I_{sppa}$ is higher for 70% DC compared to 50%. Threshold energy density is higher at higher pulse train durations. There is some variation in DC and pulse train duration effects based on PD. Pulsed sonication is more effective than continuous. | None. | CW and PW<br>Multiple parameters manipulated simultaneously. |
| *King et al., 2013*[25] | Anesthetised mice | EMG success rate, latency, duration and amplitude | $f_0$, $I_{sppa}$, $I_{spta}$, PD and PRF | EMG success rate increases with increasing $I_{sppa}$ and $I_{spta}$ (for both CW and PW). Lower $I_{sppa}$ is required to achieve the same success rates at lower $f_0$ (for CW). EMG latency decreases with increasing $I_{sppa}$ (for CW). EMG success rate increases as PD (for CW) increases. CW is more effective than PW, only at relatively high $I_{sppa}$ and $I_{spta}$. EMG success rate increases as PRF increases. Success rate is determined by a combination of pressure and PD. | Uncoupling of transducer, or sonication of cervical region near brainstem leads to reduction of EMG success rate to levels similar to sham no-US condition. | CW and PW<br>Total energy delivered increases as PRF increases. |
| *Younan* et al.*, 2013*[24] | Anesthetised rats | Visible muscle contraction or movement | Pressure | Likelihood of motor response increases as pressure increases. | None. | PW |
| *Tufail* et al.*, 2010*[6] | Anesthetised mice | EMG amplitude | $f_0$ and $I_{spta}$ | EMG amplitude decreases as $I_{spta}$ and $f_0$ increase. | Activation of isolated muscle groups elicited by shifting the transducer over the motor cortex, but not possible to generate reliable maps. | PW<br>Multiple parameters manipulated, making it difficult to isolate any effects due to $f_0$ and Ispta. |

*$I_{sppa}$: spatial peak pulse averaged intensity; $I_{spta}$: spatial peak time averaged intensity; $I_{sapa}$: spatial averaged pulse averaged intensity; $I_{sata}$: spatial averaged time averaged intensity; $f_0$: fundamental frequency; PD: pulse duration; PRF: pulse repetition frequency; DC: duty cycle; CW: continuous wave; PW: pulsed wave; MEP: motor evoked potential; BOLD: blood-oxygen-level-dependent signal; iTBUS: intermittent theta burst ultrasound stimulation; cTBUS: continuous theta burst ultrasound stimulation*

**Table 3:**
Summary of human studies that directly compare TUS parameters.

| | *Outcome measure(s)* | *Directly compared TUS parameters* | *Relevant findings* | *Shams/ controls in parameter comparison experiments* | *Notes* |
|---|---|---|---|---|---|
| *Zadeh et al. 2024*[70] | TMS evoked- MEP amplitude and latency | PRF | Decrease in MEP amplitude after 10 and 100 Hz, but not 1000 Hz. No effect of TUS on MEP latency. | Inactive sham with no ultrasound delivered. | PW<br>Offline<br>Total energy delivered is matched by decreasing PD as PRF increases. |
| *Zeng et al. 2024*[71] | TMS evoked- MEP amplitude, SICI, ICF, SICF, RMT | $I_{sppa}$, PRF, DC, and pulse train duration | Larger increase in MEP amplitude at 9.04 (2.26) compared to 4.52 (1.13) W/cm$^2$ $I_{sppa}$ (intracranial). Decrease in SICI at 9.04 but not at 4.52 W/cm$^2$ $I_{sppa}$. Increase in SICF at 9.04 but not at 4.52 W/cm$^2$ $I_{sppa}$.<br>Larger increase in MEP amplitude at 5 compared to 2 and 10 Hz PRF. Decrease in SICI at 10 but not 2 Hz PRF. Increase in SICF at 10 but not 2 Hz PRF. Effects of 5 Hz PRF relative to other PRFs not reported.<br>Larger increase in MEP amplitude at 10 and 15% compared to 5% DC.<br>Larger increase in MEP amplitude with 120 s compared to 40 and 80 s pulse train duration. Increase in SICF at 120 s but not 40 s pulse train duration. Effects of 80 s pulse train duration relative to other pulse train durations not reported. | None. | PW<br>Offline<br>Multiple parameters manipulated simultaneously. |
| *Kop et al., 2024*[10] | TMS evoked- MEP amplitude | $I_{sppa}$ and pulse train duration | $I_{sppa}$ ranging from 4.34 to 65 W/cm$^2$ did not result in direct neuromodulation of MEPs. Pulse train durations of 500 ms, but not 100 ms, reduced MEP amplitudes, but also for active control and sound-sham conditions. SImilarly, a dose-response effect of | Sound-sham, active control TUS, and inactive control TUS | PW |
| | | | confounds was observed for auditory confound volume (approximated by $I_{sppa}$) and motor inhibition. | | |
| *Fomenko et al. 2020*[72] | TMS evoked- MEP amplitude | DC, pulse train duration and PRF | This investigation did not control for auditory effects, and follow-up work suggests that the effects described below are likely a combination of neuromodulatory and auditory effects[10]. Decrease in MEP amplitude at 10 and 30%, but not 50% DC, when tested in blocks. Decrease in MEP amplitude at 10%, but not 30 and 50% DC, when tested with trials with different DCs interleaved. Decrease in MEP at pulse train durations of 0.4 and 0.5 s, but not at shorter durations. In blocked design, all tested PRFs (200, 500 and 1000 Hz) lead to decrease in MEP amplitude compared to sham. This is seen whether the total energy delivered is matched or not. In interleaved design, no decrease in MEP is observed at any PRF. | Inactive sham with no ultrasound delivered. | PW |

*$I_{sppa}$: spatial peak pulse averaged intensity; PRF: pulse repetition frequency; DC: duty cycle; PW; pulsed wave; TMS: transcranial magnetic stimulation; MEP: motor evoked potential; SICI: short-interval intracortical inhibition; ICF: intracortical facilitation; SICF: short-interval intracortical facilitation; RMT: resting motor threshold*

**Funding:** KBP is supported by grants from the National Institutes of Health (NIH R01 MH131684, NIH R01NS112152, NIH R01 EB032743). CJS holds a Senior Research Fellowship, funded by the Wellcome Trust (224430/Z/21/Z). LV is supported by a VIDI fellowship (18919) funded by the Dutch Research Council (NWO), by an Open Call HHT (HiTMaT-38H3) funded by Holland High Tech, and is a co-applicant on an EIC Pathfinder project (CITRUS, 101071008) funded by the European Innovation Council (EIC) and on an ERC Advanced project (MediCoDe) funded by the European Research Council (ERC).

**Declarations of interest:** KBP has no competing interests related to the reported work. Unrelated to the reported work, KBP has a relationship with MR Instruments, having received equipment on loan, and with Attune Neurosciences as a consultant. LV has no competing interests related to the reported work. Unrelated to the reported work, LV has a relationship with Brainbox Initiative as a member of the scientific committee, with Nudge LLC, having received consulting fees, with Sonic Concepts Ltd, having received equipment on loan, and with Image Guided Therapy, having received equipment on loan.

**References:**

1. Blackmore, J., Shrivastava, S., Sallet, J., Butler, C. R. & Cleveland, R. O. Ultrasound neuromodulation: a review of results, mechanisms and safety. *Ultrasound Med. Biol.* 45, 1509–1536 (2019).
2. Nandi, T. *et al.* Biophysical effects and neuromodulatory dose of transcranial ultrasonic stimulation. Preprint at https://arxiv.org/abs/2406.19869 (2024).
3. Murphy, K. R. *et al.* A tool for monitoring cell type–specific focused ultrasound neuromodulation and control of chronic epilepsy. *Proc. Natl. Acad. Sci.* 119, e2206828119 (2022).
4. Brinker, S. T. *et al.* Focused Ultrasound Platform for Investigating Therapeutic Neuromodulation Across the Human Hippocampus. *Ultrasound Med. Biol.* 46, 1270–1274 (2020).
5. Bubrick, E. J. *et al.* Transcranial ultrasound neuromodulation for epilepsy: A pilot safety trial. *Brain Stimulat.* 17, 7–9 (2024).
6. Tufail, Y. *et al.* Transcranial Pulsed Ultrasound Stimulates Intact Brain Circuits. *Neuron* 66, 681–694 (2010).
7. Prieto, M. L., Firouzi, K., Khuri-Yakub, B. T. & Maduke, M. Activation of Piezo1 but Not NaV1.2 Channels by Ultrasound at 43 MHz. *Ultrasound Med. Biol.* 44, 1217–1232 (2018).

8. Menz, M. D. *et al.* Radiation Force as a Physical Mechanism for Ultrasonic Neurostimulation of the *Ex Vivo* Retina. *J. Neurosci.* 39, 6251–6264 (2019).
9. Menz, M. D., Oralkan, Ö., Khuri-Yakub, P. T. & Baccus, S. A. Precise Neural Stimulation in the Retina Using Focused Ultrasound. *J. Neurosci.* 33, 4550–4560 (2013).
10. Kop, B. R. *et al.* Auditory confounds can drive online effects of transcranial ultrasonic stimulation in humans. *eLife* 12, (2024).
11. Guo, H. *et al.* Ultrasound Produces Extensive Brain Activation via a Cochlear Pathway. *Neuron* 98, 1020-1030.e4 (2018).
12. Sato, T., Shapiro, M. G. & Tsao, D. Y. Ultrasonic Neuromodulation Causes Widespread Cortical Activation via an Indirect Auditory Mechanism. *Neuron* 98, 1031-1041.e5 (2018).
13. Mohammadjavadi, M. *et al.* Elimination of peripheral auditory pathway activation does not affect motor responses from ultrasound neuromodulation. *Brain Stimulat.* 12, 901–910 (2019).
14. Yoo, S., Mittelstein, D. R., Hurt, R. C., Lacroix, J. & Shapiro, M. G. Focused ultrasound excites cortical neurons via mechanosensitive calcium accumulation and ion channel amplification. *Nat. Commun.* 13, 493 (2022).
15. Plaksin, M., Shoham, S. & Kimmel, E. Intramembrane Cavitation as a Predictive Bio-Piezoelectric Mechanism for Ultrasonic Brain Stimulation. *Phys. Rev. X* 4, 011004 (2014).
16. Plaksin, M., Kimmel, E. & Shoham, S. Cell-Type-Selective Effects of Intramembrane Cavitation as a Unifying Theoretical Framework for Ultrasonic Neuromodulation. *eneuro* 3, ENEURO.0136-15.2016 (2016).
17. Tyler, W. J. The mechanobiology of brain function. *Nat. Rev. Neurosci.* 13, 867–878 (2012).
18. Burman Ingeberg, M., Van Houten, E. & Zwanenburg, J. J. M. Estimating the viscoelastic properties of the human brain at 7 T MRI using intrinsic MRE and nonlinear inversion. *Hum. Brain Mapp.* hbm.26524 (2023).
19. Bergmann, T. O. Brain State-Dependent Brain Stimulation. *Front. Psychol.* 9, 2108 (2018).
20. Bradley, C., Nydam, A. S., Dux, P. E. & Mattingley, J. B. State-dependent effects of neural stimulation on brain function and cognition. *Nat. Rev. Neurosci.* 23, 459–475 (2022).
21. Wattiez, N. *et al.* Transcranial ultrasonic stimulation modulates single-neuron discharge in macaques performing an antisaccade task. *Brain Stimulat.* 10, 1024–1031 (2017).
22. Yang, P.-F. *et al.* Bidirectional and state-dependent modulation of brain activity by transcranial focused ultrasound in non-human primates. *Brain Stimulat.* 14, 261–272 (2021).

23. Prieto, M. L., Firouzi, K., Khuri-Yakub, B. T., Madison, D. V. & Maduke, M. Spike frequency–dependent inhibition and excitation of neural activity by high-frequency ultrasound. *J. Gen. Physiol.* 152, (2020).
24. Younan, Y. *et al.* Influence of the pressure field distribution in transcranial ultrasonic neurostimulation: Influence of pressure distribution in transcranial ultrasonic neurostimulation. *Med. Phys.* 40, 082902 (2013).
25. King, R. L., Brown, J. R., Newsome, W. T. & Pauly, K. B. Effective parameters for ultrasound-induced in vivo neurostimulation. *Ultrasound Med. Biol.* 39, 312–331 (2013).
26. Sack, I. *et al.* The impact of aging and gender on brain viscoelasticity. *NeuroImage* 46, 652–657 (2009).
27. Murphy, M. C. *et al.* Regional brain stiffness changes across the Alzheimer's disease spectrum. *NeuroImage Clin.* 10, 283–290 (2016).
28. Riis, T. & Kubanek, J. Effective Ultrasonic Stimulation in Human Peripheral Nervous System. *IEEE Trans. Biomed. Eng.* 69, 15–22 (2022).
29. Attali, D. *et al.* Three-layer model with absorption for conservative estimation of the maximum acoustic transmission coefficient through the human skull for transcranial ultrasound stimulation. *Brain Stimulat.* 16, 48–55 (2023).
30. Vion-Bailly, J., Suarez-Castellanos, I. M., Chapelon, J., Carpentier, A. & N'Djin, W. A. Neurostimulation success rate of repetitive-pulse focused ultrasound in an in vivo giant axon model: An acoustic parametric study. *Med. Phys.* 49, 682–701 (2022).
31. Ye, P. P., Brown, J. R. & Pauly, K. B. Frequency Dependence of Ultrasound Neurostimulation in the Mouse Brain. *Ultrasound Med. Biol.* 42, 1512–1530 (2016).
32. Kim, H., Chiu, A., Lee, S. D., Fischer, K. & Yoo, S.-S. Focused Ultrasound-mediated Non-invasive Brain Stimulation: Examination of Sonication Parameters. *Brain Stimulat.* 7, 748–756 (2014).
33. King, R. L., Brown, J. R., Newsome, W. T. & Pauly, K. B. Effective Parameters for Ultrasound-Induced In Vivo Neurostimulation. *Ultrasound Med. Biol.* 39, 312–331 (2013).
34. Riis, T. & Kubanek, J. Effective Ultrasonic Stimulation in Human Peripheral Nervous System. *IEEE Trans. Biomed. Eng.* 69, 15–22 (2022).
35. Tyler, W. J. *et al.* Remote Excitation of Neuronal Circuits Using Low-Intensity, Low-Frequency Ultrasound. *PLoS ONE* 3, e3511 (2008).
36. Di Ianni, T. *et al.* High-throughput ultrasound neuromodulation in awake and freely behaving rats. *Brain Stimulat.* 16, 1743–1752 (2023).
37. Zheng, A. & Schmid, S. A review of the neural basis underlying the acoustic startle response with a focus on recent developments in mammals. *Neurosci. Biobehav. Rev.* 105129 (2023).
38. Kubanek, J., Shukla, P., Das, A., Baccus, S. A. & Goodman, M. B. Ultrasound Elicits Behavioral Responses through Mechanical Effects on

Neurons and Ion Channels in a Simple Nervous System. *J. Neurosci.* 38, 3081–3091 (2018).
39. Manuel, T. J. *et al.* Ultrasound neuromodulation depends on pulse repetition frequency and can modulate inhibitory effects of TTX. *Sci. Rep.* 10, 1–10 (2020).
40. Yu, K., Niu, X., Krook-Magnuson, E. & He, B. Intrinsic functional neuron-type selectivity of transcranial focused ultrasound neuromodulation. *Nat. Commun.* 12, 2519 (2021).
41. Heffner, R., Koay, G. & Heffner, H. Audiograms of five species of rodents: implications for the evolution of hearing and the perception of pitch. *Hear. Res.* 157, 138–152 (2001).
42. Verhagen, L. *et al.* Offline impact of transcranial focused ultrasound on cortical activation in primates. *Elife* 8, e40541 (2019).
43. Folloni, D. *et al.* Manipulation of subcortical and deep cortical activity in the primate brain using transcranial focused ultrasound stimulation. *Neuron* 101, 1109–1116 (2019).
44. Fouragnan, E. F. *et al.* The macaque anterior cingulate cortex translates counterfactual choice value into actual behavioral change. *Nat. Neurosci.* 22, 797–808 (2019).
45. Dallapiazza, R. F. *et al.* Noninvasive neuromodulation and thalamic mapping with low-intensity focused ultrasound. *J. Neurosurg.* 128, 875–884 (2018).
46. Nakajima, K. *et al.* A causal role of anterior prefrontal-putamen circuit for response inhibition revealed by transcranial ultrasound stimulation in humans. *Cell Rep.* 40, 111197 (2022).
47. Zeng, K. *et al.* Induction of Human Motor Cortex Plasticity by Theta Burst Transcranial Ultrasound Stimulation. *Ann. Neurol.* 91, 238–252 (2022).
48. Yaakub, S. N. *et al.* Transcranial focused ultrasound-mediated neurochemical and functional connectivity changes in deep cortical regions in humans. *Nat. Commun.* 14, 5318 (2023).
49. Yuan, Y., Wang, Z., Liu, M. & Shoham, S. Cortical hemodynamic responses induced by low-intensity transcranial ultrasound stimulation of mouse cortex. *NeuroImage* 211, 116597 (2020).
50. Yoon, K. *et al.* Effects of sonication parameters on transcranial focused ultrasound brain stimulation in an ovine model. *PLOS ONE* 14, e0224311 (2019).
51. Kim, H. *et al.* Suppression of EEG visual-evoked potentials in rats through neuromodulatory focused ultrasound. *NeuroReport* 26, 211–215 (2015).

52. Kim, H.-J. *et al.* Long-lasting forms of plasticity through patterned ultrasound-induced brainwave entrainment. *Sci. Adv.* 10, eadk3198 (2024).
53. Vasan, A. *et al.* Ultrasound Mediated Cellular Deflection Results in Cellular Depolarization. *Adv. Sci.* 9, 2101950 (2022).
54. Duck, F. Acoustic Dose and Acoustic Dose-Rate. *Ultrasound Med. Biol.* 35, 1679–1685 (2009).
55. Aubry, J.-F. *et al.* ITRUSST Consensus on Biophysical Safety for Transcranial Ultrasonic Stimulation. Preprint at http://arxiv.org/abs/2311.05359 (2023).
56. Miniussi, C., Harris, J. A. & Ruzzoli, M. Modelling non-invasive brain stimulation in cognitive neuroscience. *Neurosci. Biobehav. Rev.* 37, 1702–1712 (2013).
57. Sorum, B., Docter, T., Panico, V., Rietmeijer, R. A. & Brohawn, S. G. *Pressure and Ultrasound Activate Mechanosensitive TRAAK $K^+$ Channels through Increased Membrane Tension*. http://biorxiv.org/lookup/doi/10.1101/2023.01.11.523644 (2023).
58. Weinreb, E. & Moses, E. Mechanistic insights into ultrasonic neurostimulation of disconnected neurons using single short pulses. *Brain Stimulat.* 15, 769–779 (2022).
59. Fan, H. *et al.* Acoustic frequency-dependent physical mechanism of sub-MHz ultrasound neurostimulation. *Jpn. J. Appl. Phys.* 61, 127001 (2022).
60. Suarez-Castellanos, I. M. *et al.* Spatio-temporal characterization of causal electrophysiological activity stimulated by single pulse focused ultrasound: an ex vivo study on hippocampal brain slices. *J. Neural Eng.* 18, 026022 (2021).
61. Qiu, Z. *et al.* The Mechanosensitive Ion Channel Piezo1 Significantly Mediates In Vitro Ultrasonic Stimulation of Neurons. *iScience* 21, 448–457 (2019).
62. Zhu, J. *et al.* The mechanosensitive ion channel Piezo1 contributes to ultrasound neuromodulation. *Proc. Natl. Acad. Sci.* 120, e2300291120 (2023).
63. Di Ianni, T. *et al.* High-throughput ultrasound neuromodulation in awake and freely behaving rats. *Brain Stimulat.* (2023).
64. Yang, P.-F. *et al.* Differential dose responses of transcranial focused ultrasound at brain regions indicate causal interactions. *Brain Stimulat.* 15, 1552–1564 (2022).
65. Kim, H.-C., Lee, W., Kowsari, K., Weisholtz, D. S. & Yoo, S.-S. Effects of focused ultrasound pulse duration on stimulating cortical and subcortical motor circuits in awake sheep. *PLOS ONE* 17, e0278865 (2022).
66. Yuan, Y. *et al.* The Effect of Low-Intensity Transcranial Ultrasound Stimulation on Neural Oscillation and Hemodynamics in the Mouse

Visual Cortex Depends on Anesthesia Level and Ultrasound Intensity. *IEEE Trans. Biomed. Eng.* 68, 1619–1626 (2021).

67. Liu, Y. *et al.* Neuromodulation Effect of Very Low Intensity Transcranial Ultrasound Stimulation on Multiple Nuclei in Rat Brain. *Front. Aging Neurosci.* 13, 656430 (2021).
68. Wang, X., Yan, J., Wang, Z., Li, X. & Yuan, Y. Neuromodulation Effects of Ultrasound Stimulation Under Different Parameters on Mouse Motor Cortex. *IEEE Trans. Biomed. Eng.* 67, 291–297 (2020).
69. Kim, E., Anguluan, E. & Kim, J. G. Monitoring cerebral hemodynamic change during transcranial ultrasound stimulation using optical intrinsic signal imaging. *Sci. Rep.* 7, 13148 (2017).
70. Zadeh, A. K. *et al.* The effect of transcranial ultrasound pulse repetition frequency on sustained inhibition in the human primary motor cortex: A double-blind, sham-controlled study. *Brain Stimulat.* 17, 476–484 (2024).
71. Zeng, K. *et al.* Effects of different sonication parameters of theta burst transcranial ultrasound stimulation on human motor cortex. *Brain Stimulat.* 17, 258–268 (2024).
72. Fomenko, A. *et al.* Systematic examination of low-intensity ultrasound parameters on human motor cortex excitability and behavior. *eLife* 9, e54497 (2020).